\documentclass[12pt]{article}
\begin{document}
\input epsf
\def\be{\begin{equation}}
\def\bea{\begin{eqnarray}}
\def\ee{\end{equation}}
\def\eea{\end{eqnarray}}
\def\d{\partial}

\begin{flushright}
OHSTPY-HEP-T-01-014\\
hep-th/0105136
\end{flushright}
\vspace{20mm}
\begin{center}
{\LARGE Metric of the  multiply wound rotating string}
\\
\vspace{20mm}
{\bf  Oleg Lunin  and  Samir D. Mathur \\}
\vspace{4mm}
Department of Physics,\\ The Ohio State University,\\ Columbus, OH 43210, USA\\
\vspace{4mm}
\end{center}
\vspace{10mm}
\begin{abstract}

We consider a string wrapped many times around a compact circle in
space, and let
this string carry a right moving wave which imparts momentum and angular
momentum to the string. The angular momentum causes the strands of the
`multiwound' string to separate and cover the surface of a torus. We
compute the
supergravity solution for this string configuration. We map this solution by
dualities to the D1-D5 system with angular momentum that has been recently
studied.  We discuss how constructing this multiwound string solution
may help us
to relate the microscopic and macroscopic pictures of black hole absorption.

\end{abstract}
\newpage

\section{Introduction.}
\renewcommand{\theequation}{1.\arabic{equation}}
\setcounter{equation}{0}

String theory has been remarkably successful in explaining  the
thermodynamic properties of black holes
     in terms of the statistical mechanics of a  unitary microscopic
system. For example one can make a black hole
in 4+1 dimensions by compactifying 5 directions of 10-d Minkowski
space, and then wrapping D5 and D1 branes
around these compact directions \cite{stromvafa}. In the microscopic
description of
the D1-D5 system at weak coupling,  an
incoming graviton is absorbed when its energy gets converted to the
energy of vibrations traveling along
compact direction where the D1 brane is wrapped \cite{dasmathur}. Even though
the size  of this compact direction is small
(say Planck length), very low energy quanta can be absorbed by the
`effective string' formed by the D1 and
D5 branes due to the possibility of `multiwinding'
\cite{dasmathurpre}: if a string is
wrapped $n$ times along a circle of length
$2\pi R$ then it can carry excitations of wavelength $2\pi R n$
rather than $2\pi R$, and thus the excitation
threshold of such a string is low if $n$ is large.  It was shown in
\cite{maldasuss} that this excitation threshold
in fact agrees with the energy of the last few quanta that are
emitted thermally from a near extremal hole;
these quanta have a wavelength much larger than the radius of the
hole (which is  itself  a macroscopic
length).

On the other hand in the `macroscopic picture' of absorption,  a low
energy quantum  travels down the throat
of the 4+1 dimensional metric and falls through the horizon. We do
not see any sign of `multiwinding'
around the compact direction, and in fact do not see the energy of
the quantum being converted  to
energy in the compact directions. The smallness of the energy
threshold for absorption just stems from the
macroscopic diameter of the hole and the long length of the throat.
To better understand the fate of quanta
falling into a black hole and to obtain a resolution of the
`information paradox' we need to find a closer link
between these two descriptions of the absorption process.

In this paper we study a string which carries momentum, but also
carries a large amount of angular
momentum. This angular momentum causes the strands of the `multiwound
string' to spread out over a
surface of macroscopic size, and we study the spacetime metric
produced by such a configuration. It is hoped
that this study will eventually help us to understand better the
relation between the microscopic and
macroscopic pictures of absorption by a black hole.

\subsection{Angular momentum and the multiwound string}

Consider Type IIB string theory on flat spacetime, and let one
spatial direction $X^9$ be compactified to a
circle of length $2\pi R$. Let a D-string be wrapped once around  this $S^1$.
Further, we put momentum excitations
moving in one direction along the string, carrying a total momentum
\be
P={n_p\over R}
\label{one}
\ee
How much angular momentum can this system have (in the transverse
space $X^1,\dots ,  X^8$)? The
unexcited D-string could itself be in one of 256 ground states, which
have spins ranging from 0 to 2.  We will
ignore this spin in what follows.  The momentum modes  are carried by
open strings moving
along the D-string. Each  open string is in  one of the  massless
ground states, and carries a momentum
$n_i/R$, with $\sum n_i=n_p$.  The maximum angular momentum is
achieved if we let each open string have
$n_i=1$, and further let it be in one of the bosonic ground states
$d_{-1/2}^\mu|0\rangle_{NS}$. Aligning all the vector
indices $\mu$ gives the state with angular momentum
\be
J_{max}=n_p=PR
\label{two}
\ee
What happens if the winding number of the D-string is $n_w$ instead
of unity? If we have $n_w$ `separately
wound' D-strings, then each open string lives on one of the
D-strings, and the value $J_{max}$ remains that
given in (\ref{two}). But the $n_w$ D-strings can join up into one
long `multiply wound' D-string of total
length
$2\pi R n_w$. Then the total momentum is still of the form
(\ref{one}), but the individual open strings
carrying this momentum have
\be
p_i={n_i\over n_w R}, ~~~\sum_i n_i = n_w n_p
\label{three}
\ee
The maximum angular momentum is again obtained when all the $n_i$
are unity (and the vector spins
are all aligned), but the value of this maximum angular momentum is
\be
J_{max}=n_wn_p=n_w PR
\label{four}
\ee

Note that when each of  the momentum modes  has such a fractional value
$1/(n_w R)$ of momentum then the
vibration they describe has a wavelength $2\pi n_w R$ along the
D-string. In executing such a vibration the
$n_w$ different strands  of the D-string  will have to move apart
from each other;
they will not move together as a single high density string on the
interval $0\le X^9\le 2\pi R$. . In this paper
we wish to study the metric produced by a D-string which is in such a
microscopic configuration. The
solution we obtain will have momentum and angular momentum related as
in (\ref{four}). From the above
argument such a configuration should necessarily exhibit a spread
among the strands of the D-string;
this is an effect that we wish to see.

In more detail, we carry out the following computations:

\medskip

(a)\quad First we look at a classical string described by the
Born-Infeld action.  If we have a purely
right--moving wave on such a string, then we show that the angular
momentum carried by the string is
bounded by the value (\ref{four}).  As a byproduct of this
computation  we note the geometry of a
string carrying the maximal allowed angular momentum: the traveling
wave on the string makes the string
profile a helix, which turns around a circle $X_1^2+X_2^2=a^2$ while
moving up in a  direction $X_9$ along
which the string is wrapped. With a large value for the winding
number $n_w$ we find that there are many
strands of the string at any given value of $z$, so that the string
covers in a dense fashion the surface of the
torus given by the product of the circle $X_1^2+X_2^2=a^2$ and the
circle in the direction $X_9$.\footnote{A related calculation using 
2-branes and 0-branes has been carried out in
\cite{mateos}.}

\medskip

(b)\quad We then make a supergravity solution that will describe such
a string configuration. This solution
will carry momentum, string winding charge and angular momentum. We
start with the metric of a neutral
black hole carrying angular momentum, and transform it by a sequence
of boosts and T-dualities so that it
has momentum and winding charges. The location of the string can be
found by looking for the points where
the dilaton goes to zero: we observe that this hypersurface has the
form of the torus mentioned in the
paragraph above.

\medskip

(c)\quad In the above method of generating the supergravity solution
we do not explicitly see the strands of
the string and how they wind around the torus. We now derive the same
supergravity solution by a
different technique: we start with the known solution of an
oscillating string, take it to have a configuration
that exhibits `multiwinding', and then smear this string uniformly
over the analog  of the circle
$X_1^2+X_2^2=a^2$ to obtain a distribution that is again uniform on
the torus.  Having derived the solution
this way, we can relate the angular momentum of the solution (found
from the metric near infinity) to the
momentum of the solution, and check that these quantities satisfy the
bound (\ref{four}).

\medskip

(d)\quad Finally, we perform a sequence of dualities to map the
winding and momentum charges of the
solution to 5-brane and 1-brane charges respectively. This maps our
solution to the one discussed recently
in \cite{bala,MaldMaoz}, where the metric was found  for the D1-D5
system with angular momentum. It was
found in these references that the configuration with maximal angular
momentum was in fact a smooth
geometry with no singularities. We find that in a certain limit of
parameters both
the dual descriptions give geometries that are flat outside a `doughnut'
shaped tube of small thickness and large length. We discuss how such tubes may
exhibit low energy collective oscillations, and the physical
significance that such
modes would have.

\medskip

The plan of this paper is as follows.
Section 2  investigates the profile for a classical string in flat
space. In section 3 we
derive the geometry of the spinning string by applying boots and
dualities to the
neutral Kerr solution. Section 4 derives the same solution starting
with the known
metric of a single string, and superposing configurations to arrive at the
multiwound string. Section 5 maps the spinning string solution to the
D1-D5 system
by dualities. Section 6 is a discussion.

\section{Classical calculation for the rotating string.}
\label{SecDBI}
\renewcommand{\theequation}{2.\arabic{equation}}
\setcounter{equation}{0}

Let us begin with a simple calculation which demonstrates that if we take a
classical  string  (described by a Born-Infeld action) and excite
vibrations on it that are purely right moving
(thus giving a BPS state), then
the ratio of angular momentum of the string to  the momentum along
the string has an upper bound.
One of the directions, $X^9$, of the 10-d flat spacetime is
compactified on a circle:
\be
X^9\sim X^9+2\pi R.
\ee
     The time direction is called $X^0$. We assume that the string is
wrapped $n_w$  times around this circle
before closing on itself.  The momentum along the string will be
called $P$. The  string will in general have an
angular  momentum in
the non--compact transverse directions  $X^i, i=1, \dots 8$.
     This angular momentum is described  by the tensor
\be
M_{ij}=\frac{1}{2}\int d\sigma(X_i {\cal P}_j-X_j {\cal P}_i).
\label{angMom}
\ee
Here  ${\cal P}_i$
is the
     momentum density conjugate to $X_i$. One can characterize the
value of angular
momentum by the following invariant:
\be\label{Jdef}
J=\sqrt{2M_{ij}M_{ij}}.
\ee

We will be interested only in a special case of BPS solutions which describe
right--moving excitations of the string. We will show that for such solutions
there is an upper bound on a ratio $J/P$:
\be
\left|\frac{J}{P}\right|\le Rn_w.
\ee

The classical Dirac--Born--Infield
(Nambu--Goto) action for the string is:
\be\label{DBIact}
S=-T\int d^2\sigma\sqrt{-\mbox{det}g},
\ee
where T is the tension of the string and $g$ is the induced metric on
the string worldsheet:
\be
g_{ab}=\eta_{\mu\nu}\frac{\partial X^\mu}{\partial \sigma^a}
     \frac{\partial X^\nu}{\partial \sigma^b}=
\left(
\begin{array}{cc}
\dot{X}^\mu \dot{X}_\mu & \dot{X}^\mu X'_\mu\\
\dot{X}^\mu X'_\mu & {X^\mu}' X'_\mu
\end{array}
\right).
\ee
Here $X'_\mu$ is the derivative of $X_\mu$ with respect to $\sigma^1=\sigma$,
while  $\dot X_\mu$
is the derivative of $X_\mu$  with respect to $\sigma^0=\tau$. In
terms of $X^\mu$
the action (\ref{DBIact}) reads:
\be
S=-T\int d^2\sigma\sqrt{\left(\dot{X}^\mu X'_\mu\right)^2-
\left(\dot{X}^\mu \dot{X}_\mu\right)\left({X^\mu}' X'_\mu\right)}.
\ee
One can use this action to evaluate the canonical momenta:
\be\label{momentum}
P_\mu=\frac{\delta S}{\delta{\dot X}^\mu}=
T\int \frac{d\sigma}{\sqrt{\left(\dot{X}^\mu X'_\mu\right)^2-
\left(\dot{X}^\mu \dot{X}_\mu\right)\left({X^\mu}' X'_\mu\right)}}
\left\{\dot{X}_\mu\left({X^\nu}' X'_\nu\right)-X'_\mu
\left(\dot{X}^\nu X'_\nu\right)\right\}.
\ee
We will need this expression later.

Let us now impose the static gauge. We want to have $0\le \sigma <
2\pi$. Since the string has winding
number $n_w$ we have
\be
X^9(\sigma+2\pi)=X^9(\sigma)+2\pi Rn_w.
\ee
Thus the static gauge condition is
\be\label{gauge}
X^0=\tau,\qquad X^9=Rn_w\sigma.
\ee
In this gauge the action reads:
\be
S=-T\int d^2\sigma\sqrt{R^2n_w^2-R^2n_w^2\dot{{\bf X}}^2+
{\bf X}'^2-{\dot{{\bf X}}}^2 {\bf X}'^2+
(\dot{{\bf X}}\cdot{\bf X}')^2}.
\label{maction}
\ee
Here $X$ is the eight dimensional vector $(X^1,\dots,X^8)$ and the
scalar product is
defined in the usual way:
\be
{\bf C}\cdot{\bf D}=\sum_{i=1}^8 C^iD^i
\ee

We are looking for  solutions that describe purely   right moving
vibrations of the
string.  These vibrations will move at the speed of light, so $\bf X$
has the form
\be\label{LMAnzatz}
{\bf X}(\tau,\sigma)={\bf X}({\tilde\sigma}),\qquad
\tilde\sigma=\sigma-{\tau\over Rn_w}.
\ee
(It can be readily verified that such an $\bf X$ satisfies the
equations of motion.)
In particular, this anzatz allows us to write the $\tau$ derivative
in terms of the
derivative with respect to $\sigma$:
\be
{\dot{\bf X}}=-{1\over Rn_w}{\bf X}'
\ee
For such excitations the square root in (\ref{maction}) simplifies to
$Rn_w$, and we get for the momentum
      (\ref{momentum}) along the direction $X^9$:
\be\label{PrefinP9}
P_9=-T\int d\sigma \dot{{\bf X}}\cdot{\bf X}'={T\over Rn_w}\int d\sigma
\left({\bf X}'\right)^2.
\ee
For the angular momentum (\ref{angMom}) in the non--compact
directions one finds:
\be
M_{ij}=-{T\over 2}\int d\sigma\left(X_iX'_j-X_jX'_i\right).
\ee

Since $
X_i({\tilde\sigma}+2\pi )=X_i({\tilde\sigma})$ , we can write
\be
X_i({\tilde\sigma})=C_i(0)+\sum_{n=1}^\infty
\left(C_i(n)\cos n{\tilde\sigma}+D_i(n)\sin n{\tilde\sigma}\right).
\ee
Substituting this expansion into (\ref{PrefinP9}), we find:
\be\label{finP9}
P_9={T\pi\over Rn_w}\sum_{i=1}^8\sum_{n=1}^\infty
n^2\left(C^2_i(n)+D^2_i(n)\right).
\ee
For the angular momentum $M_{ij}$ one gets:
\be
M_{ij}=-{T\pi\over 2}\sum_{n=1}^\infty 2n
\left[C_i(n)D_j(n)-C_j(n)D_i(n)\right],
\ee
and for $J$ (see (\ref{Jdef})):
\be
J={T\pi }\left(\sum_{m,n=1}^\infty 4mn\left\{
{\bf C}(n)\cdot{\bf C}(m)~~ {\bf D}(n)\cdot{\bf D}(m)-
{\bf C}(n)\cdot{\bf D}(m) ~~{\bf D}(n)\cdot{\bf C}(m)\right\}\right)^{1/2}
\ee
Let us now compare this expression with $P_9$:
\be
P_9={T\pi\over Rn_w}\left(\sum_{m,n=1}^\infty
m^2n^2\left({\bf C}^2(n)+{\bf D}^2(n)\right)
\left({\bf C}^2(m)+{\bf D}^2(m)\right)\right)^{1/2}
\ee
Using the inequality
\be\label{ConjBound}
({\bf a}^2+{\bf c}^2)({\bf b}^2+{\bf d}^2)\ge 4\left|
({\bf a}\cdot{\bf b})({\bf c}\cdot{\bf d})-
({\bf a}\cdot{\bf d})({\bf c}\cdot{\bf b})\right|,
\ee
we find that:
\be
|J|<|P_9 n_w R|
\ee

In the above relation we obtain equality of the two sides if and only
if the coefficients $C_i(n), D_i(n)$
satisfy two conditions:
\bigskip

(a)\quad
     Only the lowest harmonics of vibration should be excited; i.e.
${\bf C}(n)={\bf D}(n)=0$ for $n\ne1$. ($C_i(0)$ can be set to
zero by a translation.)
\bigskip

(b)\quad The coefficients of these lowest harmonics should satisfy
the relations
\be\label{mrelation}
{\bf C}(1)\cdot {\bf D}(1)=0,\qquad ({\bf C}(1))^2=({\bf D}(1))^2.
\ee
These relations imply that in the inequality (\ref{ConjBound}) we get
${\bf  a}={\bf  b}, {\bf  c}={\bf  d}$
and that ${\bf  a} \cdot {\bf d}={\bf  c}\cdot{\bf  b}=0$.

Thus we get the following geometric picture of the string which
carries the maximal possible $J$ for a given
$P$. The two vectors ${\bf C}, {\bf D}$ give a 2-dimensional plane
transverse to the direction $X^9$; without
loss of generality we can let these be the directions $X^1, X^2$.
The string describes a helix, describing a circle
in this $X^1-X^2$ plane while moving up in the direction
$X^9$.  The string closes back on itself only after $n_w$ turns
around the compact direction $X^9$, so at  any given
value of
$X^9$ we see $n_w$ evenly spaced strands of the string.  If $n_w$ is
large, then the string covers very
densely the  toroidal surface given by the circle
$(X^1)^2+(X^2)^2={\rm constant}$ times the compact
circle in $X^9$.

Note that if instead of the lowest harmonic modes $C_i(1)$, $D_i(1)$
we excite the $m$th harmonic
modes $C_i(m)$, $D_i(m)$ (while maintaining conditions
analogous to (\ref{mrelation})) then we will
get
\be\label{clJPrel}
|\frac{J}{P}|=\frac{Rn_w}{m},
\ee

We  note that the above relations for $J$ do not depend on the string
tension $T$.  The above conclusions
hold unchanged for excitations of a p-brane that are independent of
$p-1$ of the spatial directions along
the brane -- this gives an `effective string' along the remaining direction.

\section{Generating the metric.}
\label{SectGener}
\renewcommand{\theequation}{3.\arabic{equation}}
\setcounter{equation}{0}

We saw above that for a classical string in flat space, if the
winding number $n_w$ is large, and the string
carries a right moving wave with the largest possible angular
momentum, then the strands of the string
cover densely a torus given by a circle $(X^1)^2+(X^2)^2=a^2$ times
the circle $X^9$. We now want to write
down the classical metric and gauge field produced by such a string
configuration.  In the limit of large $n_w$
    the mass density of the string is essentially uniform over this
torus. There are many ways to extract the
corresponding metric from known results in the literature\footnote{In
particular we may
obtain the metric as a special case of the result in
\cite{CvY}\cite{Youm}.}, but it is easiest for our
purposes to derive this metric from first principles  
by the following method\footnote{
Such method was used for example to
generate solutions for branes and Schwarzschild black holes in 
\cite{AddGener}}.

We start with a black hole in 4+1 dimensions $x^0, x^1, \dots x^4$
carrying mass and angular momentum,
but no charge.  (We will use lower case letters for  the curvilinear
coordinates
involved in metric computations, in contrast to the upper case
letters used for the
computations in flat space. In particular note that in the metric
computations the
direction of the string is $x^5$ while in the flat space computation
it was $X^9$.) We
can regard this solution as a black brane solution  in 10 spacetime dimensions,
which is translationally invariant in the extra 5 directions
$x^5, \dots x^9$.  We boost this solution in the direction
$x^5$, thus getting a solution with mass, angular momentum and
momentum charge. We then do a
T-duality along $x^5$, so that the momentum charge becomes an
elementary string winding charge.   We
can now boost again in the $x^5$ direction, getting a solution that
now has mass, angular momentum, as well
as winding and momentum charges. Now taking the initial mass of the
hole to zero
gives an extremal BPS solution. This solution describes an elementary
string that has
winding and momentum along $x^5$, as well as angular momentum in the
transverse noncompact
directions $x^1, \dots x^4$.

Note that we have let the solution be translationally invariant in
the four directions $x^6\dots x^9$. This
invariance corresponds to taking a smearing the strings in the
directions $x^6\dots x^9$. Such a smearing
makes no essential change to the computation (the powers of
$r$ in the various metric coefficients will reflect the number of
noncompact directions, but the qualitative
physics in unchanged).  With this translational invariance we will be
able to map the solution by dualities to a
solution of D5 and D1 branes which is relevant to the 4+1 dimensional
black hole problem and which has
been studied recently  in \cite{bala,MaldMaoz}.

\bigskip

     {\bf Notation:}\quad We will be using the following notation to keep
track of dualities and their effect on
branes. T duality along the
direction $x^k$ will be denoted $T_k$, while S duality will be just
called  $S$.
If we apply the duality $T_k$ followed by $S$ then the operation will
be written as $T_kS$.
A fundamental string extending in the directions
     $t,x^k$ we will will be denoted F1$(k)$, and a D$p$ brane extending
along $t,x^{i_1},\dots,x^{i_p}$
will be called D$p(i_1,\dots,i_p)$.  The Neveu--Schwarz 5--brane
spanning directions $t,x^5,x^6,x^7,x^8,x^9$ will be denoted as
NS5$(5,6,7,8,9)$.  Momentum charge along the direction $x^k$ will be
denoted P$(k)$.

\subsection{Generating the metric  for the rotating string.}

As our starting point we take the metric for the five dimensional
rotating black
hole. In five dimensions the rotation is specified by two parameters, but we
set one of them to be zero. Thus we start from the following metric
(see for example \cite{CvY,sen}):
\bea\label{startMetric}
ds_5^2&=&-\frac{r^2+a^2\cos^2\theta-2m}{r^2+a^2\cos^2\theta}dt^2+
\frac{r^2+a^2\cos^2\theta}{r^2+a^2-2m}dr^2+(r^2+a^2\cos^2\theta)d\theta^2
\nonumber\\
&+&\left(r^2+a^2+\frac{2ma^2\sin^2\theta}{r^2+a^2\cos^2\theta}\right)
\sin^2\theta d\phi^2+r^2\cos^2\theta d\psi^2
-\frac{4ma\sin^2\theta}{r^2+a^2\cos^2\theta}dtd\phi.
\eea
Since this metric solves Einstein's equations in the vacuum, one can consider
it as a solution of low dimensional supergravity coming from either IIA or IIB
string theory. To lift the solution to ten dimensions one can add five
flat directions to the black hole (\ref{startMetric}). The resulting
metric will satisfy ten dimensional Einstein equations in the vacuum, and
thus it will also be a solution of type II supergravity with
following values of NS--NS fields:
\be\label{start10D}
ds^2=ds_5^2+\sum_{i=5}^9 dx^i dx^i,\qquad
e^{2\Phi}=1,\qquad B_{\mu\nu}=0,
\ee
while all RR fields vanish. In this paper we will always write metrics in the
string frame unless specified otherwise. In particular, the metric in
(\ref{start10D}) can be thought of as a metric in the string frame (since
the dilaton vanishes, there is no difference between the string and
Einstein frames).

We will finally be interested in the case when directions $x^5,\dots x^9$ are
compactified on a torus:
\be
x^5\sim x^5+2\pi R_5,\quad \dots \quad x^9\sim x^9+2\pi R_9.
\ee
The classical metric does not depend on whether  or not we compactify
a direction that is translationally
invariant.  Thus we can lift this solution to the covering noncompact
space, apply a boost, and then decide to
compactify the solution again along some other translationally
invariant direction.

A boost in the direction $x^5$ is parameterized by one number $z$:
\be\label{defBoost}
\left(\begin{array}{c}
t'\\{x^5}'
\end{array}\right)=
\left(\begin{array}{cc}
\cosh z &\sinh z\\
\sinh z &\cosh z
\end{array}\right)
\left(\begin{array}{c}
t\\{x^5}
\end{array}\right)
\ee
To perform the boost we just write the metric (\ref{start10D})  in
terms of $t', x'^5$, and assume that the
coordinate
$x'^5$ is  the new compact coordinate.

After this boost the  solution is characterized by three parameters:
$m$, $a$ and $z$. Since it carries
momentum charge along $x^5$ we denote it by P$(5)$.
     Ultimately we will be interested in the limit of infinite
boost, i.e. will take the mass  parameter of the metric to zero while
keeping the combination
$A=\sqrt{m}\cosh z$ fixed. This can be achieved in two different ways:
$z$ can go to either positive or negative infinity. To account for
these two possibilities
we will characterize the boost (\ref{defBoost}) by two parameters:
\be
A=\sqrt{m}\cosh z,\qquad \alpha=\mbox{sign} (z).
\ee

To prevent a proliferation of superscripts, we relabel coordinates so
that $x'^5$ is now called $x^5$. (We have
no need for the old coordinate $x^5$.)   Next we apply T-duality in
the  direction now called
$x^5$. The P$(5)$ charge changes to the  charge of fundamental string
($F1(5)$ in our notation). The solution
continues to have the   mass parameter
$m$ and the rotation parameter $a$.

Let us now apply a new boost in the $x^5$ direction:
\be\label{defBoostp}
\left(\begin{array}{c}
t'\\{x^5}'
\end{array}\right)=
\left(\begin{array}{cc}
\cosh w &\sinh w\\
\sinh w &\cosh w
\end{array}\right)
\left(\begin{array}{c}
t\\{x^5}
\end{array}\right).
\ee
As before we can characterize this boost by the alternative parameters:
\be
B=\sqrt{m}\cosh w,\qquad \beta=\mbox{sign} (w).
\ee
Since the boost does not affect the F1 charge, it only produces a new
momentum charge
$P(5)$, and the resulting solution can be labeled $[F1(5)][P(5)]$.

     Since coordinate $x^5$ plays a special role in our
construction it is convenient to introduce a new notation for it:
\be
y=x^5,
\ee

We now take the limit $m\rightarrow 0$, while keeping
$A$ and $B$ fixed. It is helpful to
introduce the functions
\be
f_A=r^2+a^2\cos^2\theta+2A^2\qquad {\mbox{and}}\qquad f_0=r^2+a^2\cos^2\theta.
\ee
(Note that
$f_0$ is just $f_A$ for $A=0$.) Then the extremal rotating metric
with charges $[F1(5)][P(5)]$ obtained after
these steps is
\bea\label{FPsolution}
{\mbox{F1(5)P(5):}}&&\nonumber\\
ds^2&=&-\frac{f_0}{f_A}(dt^2-dy^2)+\frac{2B^2}{f_A}(dt-\beta dy)^2-
\frac{4aAB}{f_A}\sin^2\theta d\phi(dt-\beta dy)\nonumber\\
&+&f_0\left(\frac{dr^2}{r^2+a^2}+d\theta^2\right)+(a^2+r^2)\sin^2\theta d\phi^2
+r^2\cos^2\theta d\psi^2+\sum_{i=6}^9 dx^i dx^i\nonumber\\
e^{2\Phi}&=&\frac{f_0}{f_A},\qquad
B_{t\phi}=-\beta B_{y\phi}=\frac{2aAB\alpha\beta\sin^2\theta}{f_A},\qquad
B_{ty}=-\frac{2A^2\alpha}{f_A}
\eea
As mentioned above this metric and other metrics written below are
string metrics unless specified
otherwise.

\section{Superposing solutions of a vibrating string}
\label{SectRotStr}
\renewcommand{\theequation}{4.\arabic{equation}}
\setcounter{equation}{0}

In the above section we derived the supergravity solution we wanted,
but we cannot see from this solution
itself the bound on the value of angular momentum that was derived
for the classical string in section 2. The
reason for can be seen from (\ref{clJPrel}).  We can see the value of
$J$ and $P$ from the classical metric, but
this ratio determines, at the microscopic level,  not the integers
$n_w$ and $m$ but   only their ratio
$n_w/m$. By taking $n_w$ large, $m$ large, we can make this ratio as
close to any real number that we want,
and thus we should find that in the classical solution even after $J$
and $P$ are fixed there is one more
parameter in the solution. This extra one parameter freedom shows up
in the parameter $a$.

Now we would like to see more geometrically the relation of the
classical solution to the integer $n_w$. To do
this we derive the classical supergravity solution  found above in a
different way. We consider the metric
of a vibrating   string, and superpose such strings to obtain a
solution with the desired symmetries. In this
process we will in fact start by choosing the values of $n_w$ and $m$
(we will take $m=1$), and so obtain a
direct map to the microscopic variables.

\subsection{The metric of a vibrating string}

The supergravity solution for a string carrying purely right moving
vibrations is well known
\cite{DGHW,CalMalPeet}.  We will follow for the most part notations
of \cite{DGHW}. In line with the calculation
of the above section, we smear the string over 4 directions $x_6\dots
x_9$; thus all quantities are
independent of these coordinates. Then the solution is
\bea\label{DGHWsolut}
ds^2&=&-e^{2\Phi}du~dv-(e^{2\Phi}-1){\dot{\vec F}} ^2dv^2+
2(e^{2\Phi}-1){\dot{\vec F}}\cdot d{\vec y}dv+d{\vec y}\cdot d{\vec y}
+\sum_{i=6}^9 dx^i dx^i,\nonumber\\
B_{uv}&=&\frac{1}{2}(e^{2\Phi}-1),\qquad
B_{vi}=\dot{F}_i(e^{2\Phi}-1),\qquad
e^{-2\Phi}=1+\frac{Q}{|{\vec y}-{\vec F}|^2}.
\eea
This solution is parameterized by the F1 charge $Q$ and four dimensional
vector ${\vec F}=(F_1,F_2,F_3,F_4)$, which is a function of only one
coordinate $v$. $\dot{F}$ denotes the derivative of $F$ with respect to this
coordinate $v$.
As one can see, the dilaton $e^{2\Phi}$ goes to zero on the surface
\be
{\vec y}={\vec F}(v),
\ee
Thus this surface is interpreted as the location of the string. If
${\vec F}=\mbox{constant}$,
the solution describes a static string, otherwise the string is oscillating.

\subsection{The chiral null models}

An interesting property of such string solutions is that we can
easily construct the solution that describes
a set of such strings instead of just one string. The wave on each
string must be carrying momentum in the
same direction, but the waveforms need not be the same on different
strings. To see how this happens, and
to construct these multi-string solutions, it is helpful to review
the chiral null models, of which the above
string solution (\ref{DGHWsolut}) is a special case.

Consider the following supergravity solution describing   a chiral
null model  \cite{HorTseytl}:
\bea
ds^2&=&H({\vec y})\left(-du~dv+K({\vec y},v)dv^2+
2A_i({\vec y},v)dy_i dv\right)+
d{\vec y}\cdot d{\vec y}+\sum_{i=6}^9 dx^i dx^i,\nonumber\\
B_{uv}&=&-G_{uv}=\frac{1}{2}H({\vec y}),\qquad
B_{vi}=-G_{vi}=-H({\vec y})A_i({\vec y},v),\qquad
e^{-2\Phi}=H^{-1}({\vec y}).\nonumber\\
\label{ChiralSolution}
\eea
Regarding $A_i$ as a gauge field we can construct the field strength
${\cal F}_{ij}=A_{j,i}-A_{i,j}$. The
functions in the chiral null model are required satisfy the equations
\be\label{solitEqn}
\partial^2 H^{-1}=0,\qquad \partial^2 K=0,\qquad \partial_i{\cal F}^{ij}=0.
\ee
Here
$\partial^2$ is  the Laplacian
in the $y_i$ coordinates. Note that the indices $i,j$ span the
subspace $\{ y_i\}$ where the metric is just
$\delta_{ij}$, and thus these indices are raised and lowered by this
flat metric.

The above solution (\ref{DGHWsolut}) for the oscillating string is a
special case of such a chiral null model. To
see this  use the gauge freedom in $B_{uv}$ to add a constant $1/2$
to the value $B_{uv}$ in
(\ref{DGHWsolut}). Then we find that the string solution
(\ref{DGHWsolut}) gives a
chiral null model  with the following choice of functions
\be\label{DablToHorow}
H^{-1}({\vec y},v)=1+\frac{Q}{|{\vec y}-{\vec F}|^2},\qquad
K({\vec y},v)=\frac{Q|\dot{\vec F}|^2}{|{\vec y}-{\vec F}|^2},\qquad
A_i({\vec y},v)=-\frac{Q\dot{F}_i}{|{\vec y}-{\vec F}|^2}.
\ee

\subsection{The FP solution (\ref{FPsolution}) as a chiral null model}

In section 3 we had obtained the solution  (\ref{FPsolution}) with
winding charge, momentum and angular
momentum  by a combination of boosts and dualities. In order to
facilitate comparison with the rotating
string solution, we write the solution (\ref{FPsolution}) as a chiral
null model also.

Introducing the
null coordinates $u=t+y$, $v=t-y$,   the metric of (\ref{FPsolution}) reads:
\bea\label{CMPmetric}
ds_{FP}^2&=&-e^{2\Phi}du~dv-
\frac{B^2}{A^2}(e^{2\Phi}-1)dv^2+
\frac{4aB}{A}(e^{2\Phi}-1)\sin^2\theta d\phi dv\nonumber\\
&+&f_0\left(\frac{dr^2}{r^2+a^2}+d\theta^2\right)+(a^2+r^2)\sin^2\theta d\phi^2
+r^2\cos^2\theta d\psi^2+\sum_{i=6}^9 dx^i dx^i
\eea
(Recall that $f_0=r^2+a^2\cos^2\theta$.)
By a change of coordinates
\be
r'=\sqrt{r^2+a^2\sin^2\theta},\qquad
\cos\theta'=\frac{r\cos\theta}{\sqrt{r^2+a^2\sin^2\theta}}, \qquad
\phi'=\phi,\qquad \psi'=\psi
\label{maone}
\ee
the last line of the metric in (\ref{CMPmetric}) can be transformed
into a flat metric:
\bea
&&f_0\left(\frac{dr^2}{r^2+a^2}+d\theta^2\right)+(a^2+r^2)\sin^2\theta d\phi^2
+r^2\cos^2\theta d\psi^2+\sum_{i=6}^9 dx^i dx^i\nonumber\\
&&\qquad=dr'^2+r'^2d\theta'^2+r'^2\sin^2\theta' d\phi'^2
+r'^2\cos^2\theta' d\psi'^2+\sum_{i=6}^9 dx^i dx^i
\label{matwo}
\eea
Introducing the Cartesian coordinates:
\bea
&&y_1=r'\sin\theta'\cos\phi',\qquad y_2=r'\sin\theta'\sin\phi',\nonumber\\
&&y_3=r'\cos\theta'\cos\psi',\qquad y_4=r'\cos\theta'\sin\psi'
\eea
and the following useful combinations:
\be
z=\sqrt{y_1^2+y_2^2},\qquad w=\sqrt{y_3^2+y_4^2}.
\ee
     the FP solution (\ref{CMPmetric}) reads:
\bea\label{FPnullCoor}
ds^2&=&e^{2\Phi}\left(-du~dv+\frac{B^2}{A^2}(e^{-2\Phi}-1)dv^2-
\frac{4aB}{A}(e^{-2\Phi}-1)\frac{y_1dy_2dv-y_2dy_1dv}{z^2+w^2+a^2+f_0}\right)
\nonumber\\
&+&d{\vec y}d{\vec y}+\sum_{i=6}^9 dx^i dx^i,\\
B_{uv}&=&-\frac{1}{2}(e^{2\Phi}-1)\sim G_{uv},\quad
B_{vi}=-G_{vi},\quad
e^{-2\Phi}=1+\frac{2A^2}{f_0}. \nonumber
\eea
The function $f_0$ is given by the following expression:
\be
f_0=\left((z^2-a^2)^2+2w^2(z^2+a^2)+w^4\right)^{1/2}.
\ee
The singularity of the solution is located on the surface where the
dilaton $e^{2\Phi}$ vanishes, i.e. on the surface $f_0=0$. On the
transverse space $y_1\dots y_4$ this surface looks like a circle:
\be
w=0,\qquad z=a.
\ee
Thus at any fixed value of time $t$ the surface where the dilaton
vanishes is a 6-dimensional spatial surface:
     four directions $x_6, \dots x_9$ over which the string is smeared,
the coordinate $y$ along the string and the direction along the circle
$y_1^2+y_2^2=a^2$.

The solution (\ref{FPnullCoor}) obtained above has a form of a chiral null
model (\ref{ChiralSolution}) with the following choice of functions:
\bea\label{myChiral}
H^{-1}=1+\frac{2A^2}{f_0},&&\qquad K=\frac{2B^2}{f_0},\nonumber\\
A_1=\frac{4ABay_2}{f_0(z^2+w^2+a^2+f_0^2)},&&
A_2=-\frac{4ABay_1}{f_0(z^2+w^2+a^2+f_0^2)},~~ A_3=A_4=0 .
\eea

\subsection{Superposing string solutions}

Let us return to the supergravity solution (\ref{DGHWsolut}) for an
oscillating string, and its representation
(\ref{DablToHorow})  as a chiral null model. We will now superpose
such string solutions to obtain the same
chiral null model (\ref{FPnullCoor}) obtained above.

This superposition will proceed in two steps. First consider a single
string wrapped in a periodic fashion
around the compact coordinate $y$. Then in the functions $H,K,A_i$
appearing in the chiral null model we will
find that for any fixed  value of $v$ we have a `single center'
solution (\ref{DablToHorow}) of the
Laplacian in the transverse space
$\{ y_i\}$.  But we could let the string be `multiwound' around the
compact direction $y$, which means that
it closes only after several turns around this direction. Such a string
can be constructed by choosing the
functions $\vec F$ in (\ref{DGHWsolut}) in the following manner
\be\label{F1F2Dabl}
F_1(v)=a\cos(\omega v+\alpha), \qquad F_2(v)=a\sin(\omega v+\alpha).
\ee
where
\be\label{quantOmega}
\omega={m\over n_w}{1\over R}
\ee
If $m/n_w$ is not an integer, then the string does not close on itself
after one cycle around the direction $y$,
but does close after a finite number of revolutions. For any given
value of $y$, we will see several strands of
the string, evenly spaced around the circle $y_1^2+y_2^2=a^2$.  Such
a configuration is a solution of the
chiral null model where the harmonic function $H^{-1}$ is a
`multi-centered' solution of the Laplace equation.

Though at this stage we have several centers of the string on the
circle $y_1^2+y_2^2=a^2$ we do not
for any finite $n_w$ have translational invariance along this circle.
To obtain a translationally invariant
solution we `smear' the multiwound string (characterized by the
integers $m, n_w$) on this circle. One can
think of this smearing operation as one where we cut up the
multiwound string into thinner strands, each
wound around $y$ in the same way as the original strand, but with the
parameter $\alpha$ in
(\ref{F1F2Dabl}) taking a different value for each of these thin
strands. In the limit where we take  a uniform
smearing, we just average the angle $\alpha$ with uniform weight over
the circle $y_1^2+y_2^2=a^2$.
Note that because of the linearity of the chiral null model we still
get an exact solution of the supergravity
equations, and that this solution reflects the value of the integers
$m, n_w$ through the value of $\omega$
which is a parameter that will appear in the smeared solution.
The smearing operation gives
\bea
\langle H^{-1}({\vec y})\rangle&=&1+\frac{Q}{2\pi}\int_0^{2\pi}
\frac{d\alpha}{(y_1-a\cos\alpha)^2+(y_2-a\sin\alpha)^2+w^2}\nonumber\\
&=&1+\frac{Q}{2\pi}\int_0^{2\pi}
\frac{d\alpha}{z^2+w^2+a^2-2za\cos(\alpha+\omega v-\phi')}\nonumber\\
&=&1+\frac{Q}{f_0},
\label{mq}
\eea
and similarly
\bea
&&\langle K({\vec y},v)\rangle=\frac{Qa^2\omega^2}{f_0},\nonumber\\
&&\langle A_1\rangle=\frac{Qa\omega}{2\pi}\int_0^{2\pi}
\frac{d\alpha\sin(\alpha+\phi')}{z^2+w^2+a^2-2za\cos\alpha}=
-\frac{2Qa^2\omega y_2}{f_0}\frac{1}{z^2+a^2+w^2+f_0},\nonumber\\
&&\langle A_2\rangle=\frac{2Qa^2\omega y_1}{f_0}\frac{1}{z^2+a^2+w^2+f_0}
\label{mqp}
\eea
Here we evaluated the integrals using the following result:
\be
\int_0^{2\pi}\frac{d\alpha\cos^n\alpha}{1+a\cos\alpha}=
\frac{2\pi}{\sqrt{1-a^2}}\left(\frac{\sqrt{1-a^2}-1}{a}\right)^n .
\ee

\subsection{Comparing the solutions (\ref{myChiral}) and
(\ref{mq})--(\ref{mqp})}

We have obtained the metric of a fundamental string carrying momentum
and angular momentum in terms
of a chiral null model, in two different ways. The solution
(\ref{myChiral}) was obtained for  the metric
obtained by boosting and dualities starting with a Kerr black hole.
This solution is characterized by the
parameters $a,A,B$. The solution (\ref{mq})--(\ref{mqp}) was obtained
by  smearing the solution of a
spinning string. This latter solution is characterized by parameters
$a, \omega, Q$. Comparing the
functions in these chiral null model descriptions we find that the
parameter $a$ is the same in the two
solutions, while the parameters $\omega, Q$ are given by
\bea
Q=2A^2\nonumber \\
\omega=-{B\over aA}
\label{mten}
\eea

     \subsection{Momentum and angular momentum of the solutions}

     Let us extract the
momentum and angular momentum of the supergravity solutions that we
have found, by looking at the
behavior of the fields at infinity. We will first find these
quantities at the classical level, and then proceed to
find the integer number of number of quanta that correspond to each   quantity.

\subsubsection{Classical charges of the solution}

The string metric of the 10-D  FP solution (\ref{FPsolution})
behaves at infinity as
\bea\label{FPasympt}
ds^2&\approx&-\left(1-\frac{2A^2}{r^2}\right)
(dt^2-dy^2)+\frac{2B^2}{r^2}(dt-dy)^2-
\frac{4aAB}{r^2}\sin^2\theta d\phi(dt-dy)\nonumber\\
&+&dr^2+r^2(d\theta^2+\sin^2\theta d\phi^2+\cos^2\theta d\psi^2)
+\sum_{i=6}^9 dx^i dx^i\\
e^{2\Phi}&\approx&1-\frac{2A^2}{r^2},\qquad
B_{t\phi}=-B_{y\phi}\approx\frac{2aAB\sin^2\theta}{r^2},\qquad
B_{ty}\approx-\frac{2A^2}{r^2}
\eea
We should really extract the conserved quantities from the Einstein
metric, but since we will be using parts
of the metric that are vanish at infinity the correction we get from
the presence of the dilaton is subleading,
and does not affect the value of the conserved quantity. (By
contrast, if we were computing the mass then we
will have to use the Einstein metric explicitly.)

The momentum is extracted from the metric coefficient $g_{ty}$. Let a
classical solution be translation
invariant in the directions $x_\perp$, and let $D$ be the number of
remaining spacetime dimensions. The
direction $y$ is one of the directions in $x_\perp$. Then if $r$ is
the radial coordinate in the $D$ dimensional
spacetime, and if near infinity
\be
g_{ty}\sim {q\over r^{D-3}}
\ee
then the momentum of the solution in the direction $y$,  per unit
volume in the space $x_\perp$,  is
\be
{\cal P}_y=-{(D-3)\Omega_{D-2} q\over 16\pi G}
\ee
where
$\Omega_{D-2}$ is the volume of the unit $D-2$ sphere and $G$ is the
Newton's constant of the entire
spacetime (the D dimensional part together with the directions in
$x_\perp$).  Using this relation we get
for our solution
\be
{\cal P}_y= {\pi B^2\over 2 G}
\ee
and the total momentum is
\be
P_y={\pi B^2\over 2 G} (2\pi R) (2\pi)^4 V
\label{mpy}
\ee
where $2\pi R$ is the length of the direction $y$ and $(2\pi)^4 V$ is
the volume in the directions $x_6, \dots
x_9$.

The angular momentum is extracted from the coefficient $g_{t\phi}$.
If near infinity
\be
g_{t\phi}\sim {\tilde q\over r^{D-3}}
\ee
then the angular momentum in the direction $\phi$ per unit volume in
$x_\perp$ is
\be
{\cal J}_\phi={\Omega_{D-2} \tilde q\over 8\pi G}
\ee
We then find that for our solution
\be
{\cal J}_\phi=-{aAB\pi\over 2G}
\ee
and the total angular momentum is
\be
J_\phi=-{aAB\pi\over 2G} (2\pi R) (2\pi)^4 V
\ee

In terms of the parameters  used in the description  (\ref{mq})--(\ref{mqp})
we find (using (\ref{mten}))
\bea
P_y={\pi a^2 Q \omega^2\over 4G}(2\pi R) (2\pi)^4 V\nonumber \\
J_\phi={\pi a^2 Q \omega\over 4G}(2\pi R) (2\pi)^4 V 
\eea
Thus
\be
{J_\phi\over P_y}={1\over \omega}
\ee
Using the value (\ref{quantOmega}) of $\omega$ from the geometric
picture of the
string, we get
\be
{J_\phi\over P_y}={n_w R\over m}
\label{mel}
\ee
in agreement with the classical result (\ref{clJPrel}) from the flat
space computation.
This agreement is of course not a surprise -- we just obtain  a check on  our
gravity calculations since we have now computed the momentum and angular
momentum from the  metric produced by the string rather than from the string
profile itself.

\subsubsection{Charges in quantized units}

In the 10-d string theory the Newton's constant is
\be
G=g^28\pi^6 \alpha'^4
\ee
where $g$ is the string coupling.
Let us set $\alpha'=1$. Then  quantum mechanically (\ref{mpy}) gives
\be
P={2B^2RV\over g^2}={n_p\over R}
\ee
Substituting this in (\ref{mel}) gives
\be
J_\phi={n_pn_w\over m}
\ee
as expected.

We can also find the number of strings from the gauge field $B_{\mu\nu}$.
To do this one needs to evaluate the
following flux over the three sphere $(\theta,\phi,\psi)$ \cite{HorMalStr}:
\be
n_w=\frac{V}{4\pi^2g^2}\int_{S^3} e^{-\Phi}*H,
\ee
where Hodge dual is taken in $6$ dimensions: $t,r,\theta,\phi,\psi,y$. The
relevant pieces of $H$ and its dual are:
\be
H_{rty}\approx \frac{4A^2}{r^3},\qquad *H_{\theta\phi\psi}\approx
\frac{r^6\sin^2\theta\cos^2\theta}{\sqrt{r^6\sin^2\theta\cos^2\theta}}H_{rty}=
4A^2\sin\theta\cos\theta.
\ee
Thus the F1 charge $n_w$ is given by:
\be
n_w=\frac{4A^2V}{4\pi^2g^2}\int d\theta d\phi d\psi |\sin\theta\cos\theta|=
\frac{2A^2V}{g^2},
\ee

\section{Relation to F1--NS5 solution.}
\renewcommand{\theequation}{5.\arabic{equation}}
\setcounter{equation}{0}

In the papers \cite{bala,MaldMaoz} an interesting
observation was made. Consider the D1-D5 system,
where the D5 branes are wrapped on the torus
composed of the compact directions $x_5, x_6, \dots
x_9$, and the D1 branes are wrapped around $x_5$.
There are two ways to have angular momentum in
this system. One way is to add momentum
excitations along $x_5$, with each momentum
mode carrying a polarization in the noncompact
directions $x_1\dots x_4$, and thus  contributing
to the  angular momentum. This kind of angular
momentum was given to the system in the
study of the 3-charge black hole \cite{vafaetal, har}. But
the D1-D5 system can also carry angular
momentum without any momentum excitations --
the D1-D5 bound state itself can have a spin that
ranges from zero to $n_1n_5$, where $n_1$ and
$n_5$ are the numbers of D1 branes and D5 branes
respectively. One way to see this fact is to perform
a set of S and T dualities (they will be listed below)
to map the D1-D5 system to a a set of elementary
strings (wound around $x_5$) carrying
momentum (along $x_5$). The latter system can, as
was discussed in the preceding sections, carry
angular momentum in the noncompact directions
$x_1, \dots x_4$; the magnitude of this angular momentum can range from
zero to $n_1n_5$.

The interesting observation of \cite{bala,MaldMaoz} was
that when the D1-D5 system (without momentum
excitations) is placed in a configuration of maximal
angular momentum $n_1n_5$, then the
corresponding classical geometry is completely
smooth, with no horizon and no naked singularity.
If the angular momentum is $\gamma n_1n_5$,
$\gamma<1$, then there is still no horizon but
there is a conical defect in the geometry at $r=0$.
Since the configuration with maximal angular
momentum is a unique one, while those with lower
angular momenta have some degeneracy, one may
speculate that the interpretation of the conical
singularity is that this singularity reflects the
nontrivial entropy of states for $\gamma<1$, and
the absence of any singularity for $\gamma=1$
reflects the uniqueness of that state.

Since the D1-D5 system  can be mapped by
dualities to the fundamental string carrying
momentum, each configuration of the D1-D5
system (possessing some value of $\gamma$) can
be mapped to some configuration of the spinning
string of the kind that we have studied in the
sections above. This leads to the natural question:
which configurations of the spinning string
are the ones dual to the smooth D1-D5 geometry
with $\gamma=1$? Can we see any analogue of this
smoothness of the latter system in the
corresponding string solution?

At first the analysis of this issue appears to pose a
puzzle. The classical spinning string geometries are
described by the following parameters. The
    string describes a helix which has a radius
$a$ and a height $R$ ($a^2=y_1^2+y_2^2$, and $R$
is the radius of the $y$ coordinate around which the
string is wrapped).  in addition we have the parameters
$A, B$ which also have units of length and which  describe the
curvature length scales of the metric due to the
string and the momentum respectively. (The
solution also has the flat torus $x_6, \dots x_9$ of
volume $(2\pi)^4 V$, but this is not very relevant
to the physics below.)

Thus there are three
dimensionless ratios that can be constructed out of
these four quantities $R,a, A, B$.  At first one might
think that some combination of these  dimensionless
ratios must map to the dimensionless parameter
$\gamma$ which describes the dual D1-D5 system.
But if such is the case, then there must be a
preferred subset of the spinning string geometries,
which map to $\gamma=1$. This appears puzzling,
since it is not clear why any of the spinning string
geometries should be more special than others.

The actual situation with the duality map is a little
more complex. As we will see below, the value of
$\gamma$ in terms of the parameters of the
spinning string is
\be
\gamma=\frac{ag^2\alpha'^3}{2RABV}
\label{mafive}
\ee
where we have temporarily restored the string
length squared parameter $\alpha'$. (In all other
relations $\alpha'$ is set to unity.) Let us set $V$ to be order
$\alpha'^2$ for
concreteness. Then we observe that if the parameters $a, R, A, B$ are all
macroscopic lengths, i.e. many times the string length $\sqrt{\alpha'}$, then
$\gamma$ is near {\it zero} if $a,R,A,B$ are all lengths of comparable order.
In other words if we hold the relative values of $a,R,A,B$ fixed and
take the classical
limit, then any such system maps to a dual D1-D5 system with $\gamma=0$.
If, on the other hand we want $\gamma=1$ then we will need to have the radius
of the helix, $a$, much larger than the scales $R, A, B$. Thus the
value $\gamma=1$
is obtained in the spinning string picture as a configuration whose
length ratios are
not classical in the sense that they diverge as the Planck constant is
taken to zero.

We will discuss the physics of such configurations further in the
discussion. In the
remainder of this section we construct explicitly the duality map
between the F1-P
system (the elementary string with momentum) and the NS5--F1 system
    (the elementary string and NS5--brane system, which
is S-dual to the D1-D5 system); the metric for the latter is given
for example in
\cite{CvY,CvLars,bala,MaldMaoz}.

First we introduce a convenient notation. Let us label the system of
two kinds of
charges  as
\be
\begin{array}{|c|}
P(5)\\F1(5)
\end{array}
\ee
Application of S duality transforms $F1(5)$ into $D1(5)$ and leaves P(5)
invariant. We will use following diagram to describe the action of such a
duality:
\be
\begin{array}{|c|c|c|}
\mbox{P(5)}&&\mbox{P(5)}\\
&\stackrel{\textstyle S}{\rightarrow}&\\
\mbox{F1(5)}&&\mbox{D1(5)}\\
\end{array}
\ee
The P(5)--F1(5) system is connected with F1--NS5 by the following chain of
dualities:
\be\label{dualMap}
\begin{array}{|c|c|c|c|c|c|c|c|c|}
\mbox{P(5)}&&\mbox{P(5)}&&\mbox{P(5)}&&\mbox{P(5)}&&\mbox{F1(5)}\\
&\stackrel{\textstyle S}{\rightarrow}&&
\stackrel{\textstyle T6789}{\longrightarrow}&&
\stackrel{\textstyle S}{\rightarrow}&&
\stackrel{\textstyle T56}{\longrightarrow}&\\
\mbox{F1(5)}&&\mbox{D1(5)}&&\mbox{D5(56789)}&&
\mbox{NS5(56789)}&&\mbox{NS5(56789)}
\end{array}
\ee
It is also important to see what happens with various parameters of the
solution (e.g. coupling constant and radii) under such dualities. In previous
sections we have introduced the string coupling  $g$, the radius $R$
along $x_5$,
and the  volume  $(2\pi)^4 V$ of the torus $T^4$ in the directions
$x_6\dots x_9$.
Let us define explicitly the radius in the direction $x_6$ through ($x^6\sim
x^6+2\pi R_6$); we will need this radius in order to do T-dualities
in $x_6$.  Note that
in our notation the charges
$A$ and $B$ are not invariant under $S$ duality, but transform in the
same way as
coordinates (see appendix for details). We find the  following
transformations for
the parameters:
\be\label{DualParam}
\left(\begin{array}{c}
g\\A\\R\\R_6\\V
\end{array}\right)
\stackrel{\textstyle S}{\rightarrow}
\left(\begin{array}{c}
1/g\\A/\sqrt{g}\\R/\sqrt{g}\\R_6/\sqrt{g}\\V/g^2
\end{array}\right)
\stackrel{\textstyle T6789}{\rightarrow}
\left(\begin{array}{c}
g/V\\A/\sqrt{g}\\R/\sqrt{g}\\\sqrt{g}/R_6\\g^2/V
\end{array}\right)
\stackrel{\textstyle S}{\rightarrow}
\left(\begin{array}{c}
V/g\\A\sqrt{V}/g\\R\sqrt{V}/g\\\sqrt{V}/R_6\\V
\end{array}\right)
\stackrel{\textstyle T56}{\rightarrow}
\left(\begin{array}{c}
R_6/R\\A\sqrt{V}/g\\g/(R\sqrt{V})\\R_6/\sqrt{V}\\R_6^2
\end{array}\right)
\ee
Transformation laws for $B$ and $a$ are the same as for $A$.

After performing these dualities on the FP solution, one gets the following
NS5-F1 (i.e. NS5--brane and fundamental string)  solution (see
appendix for the
details):
\bea\label{FNS}
ds^2&=&-\frac{{\tilde f}_0}{{\tilde f}_{B}}(dt^2-d{\tilde y}^2)+
{\tilde f}_{ A}(\frac{dr^2}{r^2+{\tilde a}^2}+d\theta^2)+
(r^2+2{\tilde A}^2-\frac{2{\tilde A}^2
{\tilde a}^2\cos^2\theta}{{\tilde f}_{ B}})
\cos^2\theta d\psi^2\nonumber\\
&+&(r^2+{\tilde a}^2+2{\tilde A}^2+\frac{2{\tilde A}^2{\tilde a}^2
\sin^2\theta}{{\tilde f}_{ B}})
\sin^2\theta d\phi^2
-\frac{4{\tilde A}{\tilde B}{\tilde a}}{{\tilde f}_{ B}}\sin^2\theta dtd\phi+
\sum_{i=6}^9 dx^i dx^i\nonumber\\
&-&\frac{4{\tilde A}{\tilde B}{\tilde a}\alpha\beta}{{\tilde f}_{ B}}
\cos^2\theta
d{\tilde y}d\psi\\
e^{2\Phi}&=&\frac{{\tilde f}_{ A}}{{\tilde f}_{ B}},\qquad
B_{ty}=-\frac{2\beta {\tilde B}^2}{{\tilde f}_{ B}}, \qquad
B_{t\psi}=-\frac{2{\tilde A}{\tilde B}{\tilde a}\alpha\cos^2\theta}
{{\tilde f}_{ B}},
\nonumber\\
\label{FNSend}
B_{y\phi}&=&-\frac{2{\tilde A}{\tilde B}{\tilde a}\beta
\sin^2\theta}{{\tilde f}_{ B}},
\qquad
B_{\phi\psi}=2{\tilde A}^2\alpha\cos^2\theta+
\frac{2{\tilde A}^2\alpha{\tilde a}^2\sin^2\theta\cos^2\theta}
{{\tilde f}_{ B}}.
\eea
\be
{\tilde f}_A=r^2+{\tilde a}^2\cos^2\theta+2{\tilde A}^2.
\ee
The coordinate $x^5$ has now been called $\tilde y$   to remind
ourselves that the
compactification radii for the original coordinate $y$ (in the FP
solution) and $\tilde
y$ are different:
\be\label{periods}
y\sim y+2\pi R,\qquad {\tilde y}\sim {\tilde y}+2\pi{\tilde R},\qquad
{\tilde R}=\frac{g}{R\sqrt{V}}.
\ee
The integer  F1 and NS5 charges
associated with this solution are given by,
\bea
n'_1&=&\frac{\tilde V}{4\pi^2{\tilde g}^2}\int_{S^3}
e^{-\Phi}*H =\frac{2{\tilde B}^2\beta V'}{{\tilde g}^2}=
\frac{2B^2\beta R^2V}{g^2}=n_P,
\nonumber\\
n'_5&=&\frac{1}{4\pi^2}\int_{S^3} H =2{\tilde A}^2\alpha=
\frac{2A^2\alpha V}{g^2}=n_w,
\label{mafour}
\eea
while the angular momentum is the same as for the FP solution.

    To compare the above NS5-F1 solution with the solution in
\cite{bala,MaldMaoz}, we reduce the system (\ref{FNS}) to six dimensions
($t,r,\theta,\phi,\psi,{\tilde y}$) and look at the resulting metric in the
Einstein frame:
\bea\label{myToCompare}
ds_E^2&=&e^{-\frac{4\Phi}{6-2}}ds_6^2=-\frac{{\tilde f}_0}{\sqrt{{\tilde f}_A
{\tilde f}_{B}}}(dt^2-d{\tilde y}^2)
-\frac{4{\tilde A}{\tilde B}{\tilde a}}{\sqrt{{\tilde f}_A {\tilde f}_{B}}}
\left(\sin^2\theta dtd\phi+\alpha\beta\cos^2\theta d{\tilde y}d\psi\right)
\nonumber\\
&+&\sqrt{{\tilde f}_A {\tilde f}_{B}}(\frac{dr^2}{r^2+a^2}+d\theta^2)
+\frac{\cos^2\theta d\psi^2}{\sqrt{{\tilde f}_A {\tilde f}_{B}}}
\left({\tilde f}_A {\tilde f}_{B}-{\tilde a}^2\cos^2\theta
({\tilde f}_A+{\tilde f}_{B}-{\tilde f}_0)\right)\nonumber\\
&+&\frac{\sin^2\theta d\phi^2}{\sqrt{{\tilde f}_A {\tilde f}_{B}}}
\left({\tilde f}_A {\tilde f}_{B}+{\tilde a}^2\sin^2\theta
({\tilde f}_A+{\tilde f}_{B}-{\tilde f}_0)\right)
\eea
Let us now set $\alpha=\beta=1$, and compare the above solution with
the solution presented in \cite{bala,MaldMaoz}. We use the notation of
\cite{MaldMaoz}
\footnote{The  solution of \cite{MaldMaoz} is
for the D5-D1
system rather than NS5-F1, but since we are looking only at the Einstein metric
there is no change under the S-duality that relates these two systems.}:
\bea\label{MaldToCompare}
d{\bar s}^2&=&-\frac{\sqrt{Q_1 Q_5}}{h}(d{\bar t}^2-d{\bar y}^2)+
\sqrt{Q_1 Q_5}hf\left(d\theta^2+\frac{d{\bar r}^2}{{\bar r}^2+\gamma^2}\right)
\nonumber\\
&-&\frac{2\gamma\sqrt{Q_1 Q_5}}{hf}\left(\cos^2\theta d{\bar y}d\psi+
\sin^2\theta d{\bar t}d\phi\right)\\
&+&h\sqrt{Q_1 Q_5}\left[
\left({\bar r}^2+\frac{\gamma^2\cos^2\theta}{h^2f^2}\right)
\cos^2\theta d\psi^2+
\left({\bar r}^2+\gamma^2-\frac{\gamma^2\sin^2\theta}{h^2f^2}\right)
\sin^2\theta d\phi^2\right],\nonumber
\eea
\be
f={\bar r}^2+\gamma^2\cos^2\theta,\qquad
h=\frac{1}{{\bar R}^2f}(Q_5f+{\bar R}^2)^{1/2}
(Q_1f+{\bar R}^2)^{1/2}
\ee

The metrics (\ref{myToCompare}) and (\ref{MaldToCompare}) are related by the
change of variables
\be
{\bar R}={\tilde R},\qquad Q_1=2{\tilde B}^2,\qquad Q_5=2{\tilde A}^2,
\qquad \gamma=\frac{{\tilde a}{\tilde R}}{2{\tilde A}{\tilde B}}
\ee
\be
{\bar r}=\frac{r{\tilde R}}{2{\tilde A}{\tilde B}},\qquad
{\bar t}=\frac{t}{{\tilde R}},\qquad {\bar y}=\frac{\tilde y}{\tilde R}.
\ee
One can also rewrite the expressions for the parameters of the solution
(\ref{MaldToCompare}) in
terms of parameters of the FP solution (\ref{FPsolution}) which was
obtained for the
spinning string:
\be
{\bar R}=\frac{g}{R\sqrt{V}},\qquad Q_1=\frac{2B^2V}{g^2},\qquad
Q_5=\frac{2A^2V}{g^2},\qquad
\gamma=\frac{ag^2}{2RABV}=\frac{J}{n_1 n_P}\nonumber
\ee

\section{Discussion}
\renewcommand{\theequation}{6.\arabic{equation}}
\setcounter{equation}{0}

We have found the supergravity solution which describes a string that is wound
several times around a compact direction and carries a right moving wave of the
form (\ref{F1F2Dabl}). This solution was obtained by applying boosts
and T-dualities
to the neutral Kerr solution, and also by superposing the known solution of the
oscillating string. The momentum and angular momentum of the solution were read
off from the metric, and related to the corresponding values of these
quantities
expected from a classical computation in flat space. The metric of
the string with
momentum is dual to a D1-D5 system; this duality map was constructed
explicitly.

While there is nothing very novel about finding such supergravity solutions,
there are several reasons why we would like to study the properties
of this solution
in some detail. If we study absorption by a D1-D5 system at weak coupling, the
energy of an incoming quantum gets converted to the energy of
vibrations that run
up and down the `effective string' formed by the 1-brane - 5-brane
system.  If we
increase the coupling to obtain  a black hole geometry, then such a
model may not
hold -- the left and right moving modes may not remain weakly interacting.  One
can then try to describe the system by a dual
$AdS_3\times S^3\times M_4$ geometry \cite{malda}, but in the latter
picture we do
not, so far,  have  an idea of how an incoming quantum actually returns back to
asymptotic infinity as (information carrying) Hawking radiation.

    The consideration of angular momentum may provide a domain where
some aspects of both descriptions survive.  Consider the string with momentum
charge and large angular momentum.  Let us take the size
$R$ of the circle on which the string is wrapped to be of order the
string length, and
take the compact torus $x_6\dots x_9$ to be of string size as well.
If the angular
momentum is of order the  maximal allowed value
$J_{\max}=n_wn_p$, then using (\ref{mafive}) we see that for  any
given $g$, as $n_p, n_w$
become large, the
length scale
$a$ becomes much larger than the length scales
$A,B$ (which describe the curvature length scales of the metric due
to the winding
and momentum charges respectively).
   From the geometry (\ref{FPsolution})
we see that  the
torus on which the string is wrapped becomes very small in one
direction (the direction of
length $R$) and very long in its other direction (the circle of radius
$a$). The string location is characterized by the
region where the dilaton becomes small, and this can be seen to be $r\approx 0,
~\theta\approx \pi/2$. Away from this region, the first line of the metric in
(\ref{FPsolution}) is just $-(dt^2-dy^2)$, while the second line is
always a flat space
metric as shown by the relations (\ref{maone})--(\ref{matwo}).  Thus
we see that we obtain a  thin and long  doughnut shaped  `tube',
sitting in a spacetime that
is otherwise approximately flat.
The thickness
of the tube can be characterized by
${ max} \{A,B\}$, which describes how far the gravitational field
extends from the axis
of the tube.
    The circumference of this tube is long:
$2\pi a$. In the
coordinates (\ref{maone}) this tube is located at
\be
r'\approx a, ~~\theta\approx \pi/2,~~ 0\le\phi<2\pi.
\label{mathree}
\ee

What will be the description of low energy excitations of this multiply wrapped
string? Imagine a vibration of the string in the compact direction $x_6$. The
wavelength of the vibration is order $2\pi n_wR$, so the deformation
$\delta x_6$
does not change significantly as we go once around the circle in the
direction $y$:
$y\rightarrow y+2\pi R$. But it does change as we go halfway around
the string, i.e.
increase $\phi$ by an amount of order unity.  So $\delta x_6$ is
essentially only a
function of $\phi$ and not a function of $y$.  Thus we see that long wavelength
vibrations of the string look like low frequency vibrations of the
effective `tube'
(\ref{mathree}).  We may thus write an effective action for this
tube, and investigate
its oscillations. We will carry out such an investigation elsewhere,
but here we note
that if this tube has long lived oscillations that are superpositions
of left and right
moving modes, then we would have a situation where at strong coupling
(i.e. in a
supergravity solution) we have obtained the non-BPS state describing
both left and
right vibrations of a string.  For this string then we would have a common
description of  (non-BPS) excitations both in the weak coupling (flat
space) and
strong coupling (supergravity) domains.

Let us turn now the smooth metric found in \cite{bala,MaldMaoz} to describe
the D5-D1 system at maximum angular
momentum ($\gamma=1$). This metric has three essential length scales: the
curvature scales $\tilde A, \tilde B$ which arise from the D5 and D1 charges
respectively, and the length $\bar R$ which is the length of the
$\tilde y$ circle
at infinity. We set $\tilde A=\tilde B$, and look at the ratio $\bar
R/\tilde A$. When
$\bar R/\tilde A\gg 1$ the geometry has a large region of the form
$AdS_3\times S^3$;
some energy scales for this geometry were investigated in
\cite{mathur}. Here we
wish to consider the opposite limit: $\bar R/\tilde A\ll 1$. In this
case it can be checked
that the geometry of \cite{bala,MaldMaoz}, given in (\ref{MaldToCompare}),
is essentially
flat space outside a `tube' that has the shape of a thin doughnut.
The thickness of the
tube is $\sim \tilde A$ and the length of the tube is $2\pi a$.
Mapping this geometry
by the dualities (\ref{DualParam})  to the oscillating string FP
geometry, we find
again that the string is wrapped on the surface of a similar `doughnut', and
spacetime is essentially flat outside the  doughnut. The only
difference between the
two pictures is that in the D5-D1 geometry the spacetime inside the doughnut is
smooth, while in the FP solution the microscopic description of the
doughnut is in
terms of closely spaced strands of an elementary string. (Note however that the
smoothness of the spacetime in the former case is only a result at the level of
classical supergravity: if there was a residual conical defect of
order $1/n_5$  as
$\gamma\rightarrow 1$ then the classical limit would still indicate a smooth
spacetime at $\gamma=1$.) It would be helpful to investigate further
the low energy
excitations of
the D1-D5 system in this limit by looking at correlators at the orbifold point
of the D1-D5 CFT \cite{ml}.

\newpage

\appendix
\section{Dualities in string theory: conventions and notations.}
\renewcommand{\theequation}{A.\arabic{equation}}
\setcounter{equation}{0}

In section \ref{SectGener} we started from fundamental string with some
momentum along it and applied a chain of string dualities in order to
construct the NS5--F1 system. In this appendix we summarize the rules
describing the action of various dualities on matter fields. We use the same
conventions as \cite{johnson}.

We begin with the actions for type II supergravities, which fix normalization
of the fields. The bosonic part of the action for type IIA theory is:
\bea
S_{IIA}&=&\frac{1}{(2\pi)^7{\alpha'}^4}\int d^{10}x\sqrt{-G}\left\{
e^{-2\Phi}\left[R+4(\nabla\phi)^2-\frac{1}{12}\left(H^{(3)}\right)^2\right]
\right.\nonumber\\
&-&\left.\frac{1}{4}\left(G^{(2)}\right)^2-\frac{1}{48}\left(G^{(4)}\right)^2
\right\}
-\frac{1}{2(2\pi)^7{\alpha'}^4}\int B^{(2)}dC^{(3)}dC^{(3)}.
\eea
The field strengths of NS field $B$ and RR fields $C$ are given by:
\be
H^{(3)}=dB^{(2)},\qquad G^{(2)}=dC^{(1)},\qquad G^{(4)}=dC^{(3)}+
H^{(3)}\wedge C^{(1)}.
\ee
For the type IIB theory we have:
\bea
S_{IIB}&=&\frac{1}{(2\pi)^7{\alpha'}^4}\int d^{10}x\sqrt{-G}\left\{
e^{-2\Phi}\left[R+4(\nabla\phi)^2-\frac{1}{12}\left(H^{(3)}\right)^2\right]
\right.\nonumber\\
&-&\left.\frac{1}{12}\left(G^{(3)}+C^{(0)}H^{(3)}\right)^2-
\frac{1}{2}\left(dC^{(0)}\right)^2-
\frac{1}{480}\left(G^{(5)}\right)^2\right\}\nonumber\\
&+&\frac{1}{2(2\pi)^7{\alpha'}^4}\int \left(C^{(4)}+\frac{1}{2}B^{(2)}C^{(2)}
\right)
G^{(3)}H^{(3)}.
\eea
In this case the RR field strengths are defined by:
\be
G^{(3)}=dC^{(2)},\qquad G^{(5)}=dC^{(4)}+H^{(3)}\wedge C^{(2)}.
\ee
Here $G$  is the  string metric.
The Einstein metric is defined through
\be\label{StrToEinst}
G^E_{\mu\nu}=e^{(\Phi_0-\Phi)/2}G_{\mu\nu}=\sqrt{g}e^{-\Phi/2}G_{\mu\nu}.
\ee
(This gives the 10-d Newton's constant as
$16\pi G_N=(2\pi)^7{\alpha'}^4g^2$.)
Here $\Phi_0$ is the asymptotic value of the dilaton.
We will also need a generalization of (\ref{StrToEinst}) for  toroidal
compactifications. We assume that the ten dimensional space is a
product $M^D\times T^{10-D}$, and all fields are independent of the coordinates
on the torus \cite{MahSchw}.  The D-dimensional dilaton is given by:
\be
\phi=\Phi-\frac{1}{2}\log {\bar G},
\ee
where ${\bar G}$ is the part of the metric $G$ describing internal space
$T^{10-D}$. Then the $D$ dimensional metric in the Einstein frame is given by:
\be
G^E_{\mu\nu}=e^{-4(\phi-\phi_0)/(D-2)}G_{\mu\nu},\qquad \mu,\nu=0,\dots,D-1.
\ee

Let us now address   dualities. We begin with type IIB theory,
which is invariant under S duality. We will have the axion field
$C^{(0)}$ vanishing
in our solutions; in that case under S-duality
\be
\Phi\leftrightarrow -\Phi,\qquad
B^{(2)}_{\mu\nu}\leftrightarrow C^{(2)}_{\mu\nu},
\ee
while all other RR fields as well as the metric in the {\it Einstein frame}
stay the same.

T duality, on the other hand, transforms IIA and IIB theories into each other.
Let us assume that the space--like direction $y$ is compactified on a circle
of length $2\pi\sqrt{\alpha'} R$, then T duality transforms this coordinate
into $y'$ which now has identification $y'\sim y'+2\pi\sqrt{\alpha'}/R$. The
transformations of the NS fields are given by:
\bea
G'_{yy}=\frac{1}{G_{yy}},\qquad e^{2\Phi'}=\frac{e^{2\Phi}}{G_{yy}},&&\
G'_{\mu y}=\frac{B_{\mu y}}{G_{yy}},\qquad B'_{\mu y}=\frac{G_{\mu y}}{G_{yy}}
\nonumber\\
G'_{\mu \nu}=G_{\mu \nu}-\frac{G_{\mu y}G_{\nu y}-B_{\mu y}B_{\nu y}}{G_{yy}},
&&\
B'_{\mu \nu}=B_{\mu \nu}-\frac{B_{\mu y}G_{\nu y}-G_{\mu y}B_{\nu y}}{G_{yy}},
\eea
while for the RR potentials we have:
\bea\label{RRTDual1}
{C'}^{(n)}_{\mu\dots\nu\alpha y}&=&C^{(n-1)}_{\mu\dots\nu\alpha}-
(n-1)\frac{C^{(n-1)}_{[\mu\dots\nu|y}G_{|\alpha]y}}{G_{yy}},\\
{C'}^{(n)}_{\mu\dots\nu\alpha\beta}&=&C^{(n+1)}_{\mu\dots\nu\alpha\beta y}
+nC^{(n-1)}_{[\mu\dots\nu\alpha}G_{\beta]y}
+n(n-1)\frac{C^{(n-1)}_{[\mu\dots\nu|y}B_{|\alpha |y}G_{|\beta ]y}}{G_{yy}}.
\eea
We are interested in a special sequence of dualities (\ref{dualMap}). By
looking
at that equation, one can see that the RR fields should be transformed under
T dualities only at one step: when we apply T$6789$ to D1 brane. Since we are
starting from $C^{(1)}$ field pointing in $x^5$ direction, the only relevant
transformation is (\ref{RRTDual1}), and we will end up with RR 6--field. Such
field, however, does not enter the action for IIB theory directly, but it can
be transformed into the two form $C^{(2)}$ by using electric--magnetic duality.
Such duality simply means that the same configuration can be described by
either $p$--form $C^{(p)}$ or $(8-p)$--form $C^{(8-p)}$:
\be\label{EMdual}
dC^{(p)}=*dC^{(8-p)},
\ee
where Hodge dual is taken with respect to string metric. In particular, the
four form in type IIB theory is self--dual:
\be
dC^{(4)}=*dC^{(4)}.
\ee
In component form we have following relation between the field strengths:
\be
G^{(p+1)\mu_1\dots\mu_{p+1}}=
\frac{\epsilon^{\mu_1\dots\mu_{p+1}\nu_1\dots \nu_{9-p}}}{(9-p)!\sqrt{-G}}
G^{(8-p)}_{\nu_1\dots\nu_{9-p}},
\ee
and we normalize $\epsilon$ by $\epsilon^{01\dots 9}=1$.

While doing an S-duality we would like to have  the
metric  remain $\eta_{\mu\nu}$ at infinity. This can be achieved by doing a
rescaling of coordinates after the duality. Then the transformation rules for
S-duality are
\be
S:\qquad g'=\frac{1}{g},\qquad R'=\frac{R}{\sqrt{g}},\qquad
A'=\frac{A}{\sqrt{g}},\qquad  a'=\frac{a}{\sqrt{g}}.
\ee

For T duality in a direction $y$ with identification
$y\sim y+2\pi\sqrt{\alpha'} R_y$  we have following
transformation law:
\be
T:\qquad R'_y=\frac{1}{R_y},\qquad g'=\frac{g}{R_y}.
\ee

\section{Boosts, dualities and metrics.}
\renewcommand{\theequation}{B.\arabic{equation}}
\setcounter{equation}{0}

In this appendix we briefly describe some details of calculations omitted in
section \ref{SectGener}.  In the process we will be able to write
down explicitly the
metrics  for various 2-charge solutions (with angular momentum) that
are encountered in the course of the duality maps -- these solutions
are written
explicitly since they could be useful elsewhere.
    We will start from the
solution (\ref{FPsolution})  which describes fundamental string with momentum
propagating along it, and  apply the chain of dualities
(\ref{dualMap}) to this solution.
By applying S duality to (\ref{FPsolution}) we get:
\bea\label{D1PSolut}
{\mbox{D1--P:}}&&\nonumber\\
ds^2&=&-\sqrt{\frac{f_0}{f_A}}(dt^2-dy^2)+
\frac{2B^2}{\sqrt{f_0f_A}}(dt-\beta dy)^2-
\frac{4aAB}{\sqrt{f_0f_A}}\sin^2\theta d\phi(dt-\beta dy)\nonumber\\
&+&\sqrt{f_0f_A}\left(\frac{dr^2}{r^2+a^2}+d\theta^2\right)+
\sqrt{\frac{f_A}{f_0}}(a^2+r^2)\sin^2\theta d\phi^2
+\sqrt{\frac{f_A}{f_0}}r^2\cos^2\theta d\psi^2\nonumber\\
&+&\sqrt{\frac{f_A}{f_0}}\sum_{i=6}^9 dx^i dx^i\\
e^{2\Phi}&=&\frac{f_A}{f_0},\qquad
C^{(2)}_{t\phi}=-\beta C^{(2)}_{y\phi}=
\frac{2aAB\alpha\beta\sin^2\theta}{f_A},\qquad
C^{(2)}_{ty}=-\frac{2A^2\alpha}{f_A},
\eea
Note that as the result of this S duality, the coordinates and parameters
are rescaled according to rules represented by (\ref{DualParam}).
For example,
the relation between parameters $A,B,a,R,V,g$ involved in the D1--P solution
(\ref{D1PSolut}) and their counterparts for the FP solution, is given by:
\be
g_{D1P}=\frac{1}{g_{FP}},\qquad A_{D1P}=\frac{A_{FP}}{\sqrt{g_{FP}}},\qquad
B_{D1P}=\frac{B_{FP}}{\sqrt{g_{FP}}},
\qquad a_{D1P}=\frac{a_{FP}}{\sqrt{g_{FP}}},
\ee
\be
R_{D1P}=\frac{R_{FP}}{\sqrt{g_{FP}}},\qquad
V_{D1P}=\frac{V_{FP}}{g_{FP}^2}
\ee
It is the transformed parameters like $A_{D1P}$ that appear in the  metric
(\ref{D1PSolut}) above, but for simplicity we omit the subscripts and
write $A$ in
place of
$A_{D1P}$. We will follow this convention in all the relations that follow.

Solution (\ref{D1PSolut}) describes the system of $n_{D1}$ D1--branes with
$n_P$ units of momentum and angular momentum $J$:
\be
n_{D1}=\frac{2A^2_{D1P}V_{D1P}\alpha}{g_{D1P}},\qquad
n_{P}=\frac{2B^2_{D1P}V_{D1P}R^2_{D1P}\beta}{g^2_{D1P}},
\ee
\be
J=\frac{2a_{D1P}A_{D1P}B_{D1P}R_{D1P}V_{D1P}}{g^2_{D1P}}
\ee

At the next step we apply a sequence of T dualities (T$6789$) to transform the
D1 brane into the D5 brane. The result for the NS fields reads:
\bea\label{D5PSolut}
{\mbox{D5--P:}}&&\nonumber\\
ds^2&=&-\sqrt{\frac{f_0}{f_A}}(dt^2-dy^2)+
\frac{2B^2}{\sqrt{f_0f_A}}(dt-\beta dy)^2-
\frac{4aAB}{\sqrt{f_0f_A}}\sin^2\theta d\phi(dt-\beta dy)\nonumber\\
&+&\sqrt{f_0f_A}\left(\frac{dr^2}{r^2+a^2}+d\theta^2\right)+
\sqrt{\frac{f_A}{f_0}}(a^2+r^2)\sin^2\theta d\phi^2
+\sqrt{\frac{f_A}{f_0}}r^2\cos^2\theta d\psi^2\nonumber\\
&+&\sqrt{\frac{f_0}{f_A}}\sum_{i=6}^9 dx^i dx^i,\qquad\quad
e^{2\Phi}=\frac{f_0}{f_A}.
\eea
The relation between parameters
$A_{D5P},B_{D5P},a_{D5P},R_{D5P},V_{D5P},g_{D5P}$ entering this solution and
the parameters of (\ref{D1PSolut}) can be found using (\ref{DualParam}).
Solution (\ref{D5PSolut}) describes the system of $n_{D5}$ D5 branes with $n_P$
units of momentum and angular momentum $J$:
\be
n_{D5}=\frac{2A^2_{D5P}\alpha}{g_{D5P}},\qquad
n_{P}=\frac{2B^2_{D5P}V_{D5P}R^2_{D5P}\beta}{g^2_{D5P}},
\ee
\be
J=\frac{2a_{D5P}A_{D5P}B_{D5P}R_{D5P}V_{D5P}}{g^2_{D5P}}
\ee

Application of each of the four T dualities increases the rank of the
Ramond--Ramond fields $C$ by one. Thus the D5--P system has a RR six-form
field (which is consistent with having a D5 brane) with following nontrivial
components:
\be
C^{(6)}_{t\phi 6789}=\beta C^{(6)}_{\phi y6789}=
\frac{2ABa\alpha\beta\sin^2\theta}{f_A},\qquad
C^{(6)}_{ty6789}=-\frac{2\alpha A^2}{f_A}.
\ee
However, in order to perform S duality on this solution, it is convenient to
dualize above six-form field into the RR two-form field. This can be
done using the
general prescription of electric--magnetic duality (\ref{EMdual}). The
nontrivial components of field strength for $C^{(6)}$ are:
\bea
G^{(7)}_{tr\phi 6789}&=&-\beta G^{(7)}_{r\phi y6789}=
\frac{4rABa\alpha\beta\sin^2\theta}{f^2_A}\nonumber\\
G^{(7)}_{t\theta\phi 6789}&=&-\beta G^{(7)}_{\theta\phi y6789}=
-\frac{2ABa\alpha\beta\sin 2\theta}{f^2_A}(r^2+a^2+2A^2)
\\
G^{(7)}_{try6789}&=&-\frac{4\alpha A^2r}{f^2_A},\qquad
G^{(7)}_{t\theta y6789}=\frac{2\alpha A^2a^2\sin 2\theta}{f^2_A}
\eea
The field strength for the corresponding two--form is:
\bea
G^{(3)}_{\theta\phi\psi}&=&-\frac{2\alpha A^2r^2(a^2+r^2)
\sin 2\theta}{f_0^2},\qquad
G^{(3)}_{r\phi\psi}=-\frac{\alpha a^2A^2r\sin^2 2\theta}{f_0^2}\nonumber\\
G^{(3)}_{tr\psi}&=&-\frac{4ABa\alpha r\cos^2\theta}{f_0^2},\qquad
G^{(3)}_{t\theta\psi}=-\frac{4ABa\alpha r^2\sin\theta\cos\theta}
{f_0^2},\nonumber\\
G^{(3)}_{r\psi y}&=&\frac{4ABa\alpha\beta r\cos^2\theta}{f_0^2},\qquad
G^{(3)}_{\theta\psi y}=\frac{4ABa\alpha\beta r^2\sin\theta\cos\theta}{f_0^2}
\eea
and the nontrivial components of the RR 2-form are:
$$
C^{(2)}_{t\psi}=-\frac{2aAB\alpha\cos^2\theta}{f_0},\qquad
C^{(2)}_{\phi\psi}=\frac{2A^2\alpha(r^2+a^2)\cos^2\theta}{f_0},
$$
\be
C^{(2)}_{\psi y}=-\frac{2aAB\alpha\beta\cos^2\theta}{f_0}
\ee
Application of S duality transforms this RR two form into the NS two--form
field $B_{\mu\nu}$ and the metric and the dilaton of the transformed solution
are given by:
\bea\label{NS5PSolut}
{\mbox{NS5--P:}}&&\nonumber\\
ds^2&=&-(dt^2-dy^2)+
\frac{2B^2}{f_0}(dt-\beta dy)^2-
\frac{4aAB}{f_0}\sin^2\theta d\phi(dt-\beta dy)\nonumber\\
&+&f_A\left(\frac{dr^2}{r^2+a^2}+d\theta^2\right)+
\frac{f_A}{f_0}(a^2+r^2)\sin^2\theta d\phi^2
+\frac{f_A}{f_0}r^2\cos^2\theta d\psi^2\nonumber\\
&+&\sum_{i=6}^9 dx^i dx^i,\qquad\quad
e^{2\Phi}=\frac{f_A}{f_0}.
\eea
This solution has following charges:
\be
n_{NS5}=2A^2_{NS5P}\alpha,\qquad
n_{P}=\frac{2B^2_{NS5P}V_{NS5P}R^2_{NS5P}\beta}{g^2_{NS5P}},
\ee
\be
J=\frac{2a_{NS5P}A_{NS5P}B_{NS5P}R_{NS5P}V_{NS5P}}{g^2_{NS5P}}
\ee

Performing two T dualities (T56) on the solution (\ref{NS5PSolut}), we
finally get the F1--NS5
system (\ref{FNS}--\ref{FNSend}).

\end{document}